\documentclass[preprint]{aastex}%
\usepackage{emulateapj5}
\usepackage{apjfonts}
\def\lsim{\mathrel{\rlap{\lower4pt\hbox{\hskip1pt$\sim$}}
    \raise1pt\hbox{$<$}}}                
\def\gsim{\mathrel{\rlap{\lower4pt\hbox{\hskip1pt$\sim$}}
    \raise1pt\hbox{$>$}}}                

\def\be{\begin{equation}}
  \def\ee{\end{equation}}
\def\bea{\begin{eqnarray}}
  \def\eea{\end{eqnarray}}

\shorttitle{STATISTICAL ISOTROPY in {\it WMAP}}
\shortauthors{Hajian, Souradeep and Cornish}

\begin{document}

\title{STATISTICAL ISOTROPY OF THE {\it WILKINSON MICROWAVE ANISOTROPY PROBE} DATA: \\ A BIPOLAR POWER SPECTRUM ANALYSIS}

\author{{\large A}MIR {\large H}AJIAN {\large A}ND {\large T}ARUN
  {\large S}OURADEEP} \affil{Inter-University Centre for
  Astronomy and Astrophysics, Post Bag 4, Ganeshkhind, Pune 411007,
  India; amir@iucaa.ernet.in, tarun@iucaa.ernet.in}
\author{{\large A}ND}
\author{{\large N}EIL {\large C}ORNISH} \affil{ Department of Physics,
  Montana State University,
  Bozeman, MT 59717-3840; cornish@physics.montana.edu}  

\begin{abstract}
  
  The statistical expectation values of the temperature fluctuations of
  the cosmic microwave background (CMB) are assumed to be preserved
  under rotations of the sky. We  use the  bipolar power s  pectrum 
  (BiPS) proposed in Hajian \& Souradeep to investigate the  statistical
    isotropy (SI) of the CMB anisotropy maps recently measured by the
  {\it Wilkinson Microwave Anisotropy Probe} ({\it WMAP}). The method can isolate 
  and probe specific regions of choice in multipole space using appropriate window
  functions. The BiPS is estimated for full sky CMB anisotropy maps
  based on the first year {\it WMAP} data using a range of window
  functions. The BiPS spectra computed for both full sky maps for all
  our window functions are consistent with zero, roughly within
  $2~\sigma$.  The null BiPS results may be interpreted as an absence of
  strong violation of statistical isotropy in the first-year {\it WMAP} data
  on angular scales larger than that corresponding to $l\sim
  60$. However, pending a careful direct comparison, our results do not
  necessarily conflict with the specific SI related anomalies reported
  using other statistical tests.
  
\end{abstract}


\section{Introduction}
The cosmic microwave background (CMB) anisotropy is a very powerful
observational probe of cosmology. The recent {\it Wilkinson
Microwave Anisotropy Probe} ({\it WMAP})  data have provided a precise determination of
standard cosmological parameters reaffirming an emerging concordance
model of cosmology, structure formation \cite{wmap, sper_wmap} and, even
inflation and the early universe. 
 This remarkable success
story is almost entirely based on the measurements of the angular
power spectrum, $C_l$, of the CMB anisotropy.

In standard cosmology, the CMB anisotropy $\Delta T(\hat n)$ is
expected to be  statistically isotropic, i.e., the statistical
expectation values of $\Delta T(\hat n)$ are preserved under rotations
of the sky.  The issue of statistical isotropy (SI) applies to both 
 Gaussian and non-Gaussian CMB sky maps. Here we limit our study to the Gaussian CMB anisotropy field, where the two-point correlation
function contains all the statistical information encoded in the
field.  The importance of SI is based on a
very well known fact: if the two point correlation is rotationally invariant, one can equivalently completely specify CMB anisotropy in terms of the
Legendre transform, $C_l$, the widely used angular power spectrum of
CMB. When the SI holds, the information of the full CMB data set can
be compressed into a multipole spectrum.

After the recent release of the first year of {\it WMAP} data, the SI
of the CMB anisotropy has attracted considerable attention.  Tantalizing
evidence of SI breakdown (albeit, in very different guises) has
mounted in the {\it WMAP} first year sky maps, using a variety of
different statistics. It was pointed out that the suppression of power
in the quadrupole and octopole are aligned \cite{maxwmap}.  Further
``multipole-vector'' directions associated with these multipoles (and
some other low multipoles as well) appear to be anomalously
correlated \cite{cop04, schw04}. There are indications of asymmetry in
the power spectrum at low multipoles in opposite
hemispheres \cite{erik04a, han04, nas04}. Possibly related, are the results
of tests of Gaussianity that show asymmetry in the amplitude of the
measured genus amplitude (at about $2$ to $3\sigma$ significance)
between the north and south galactic hemispheres \cite{par04, erik04b}. Analysis of the distribution of extrema in {\it WMAP} sky maps
has indicated non-gaussianity, and to some extent, violation of
SI \cite{lar_wan04}. However, what is missing is a
common, well defined, mathematical language to quantify SI (as distinct from non Gaussianity) and the ability to ascribe
statistical significance to the anomalies unambiguously.

These results provide considerably increased motivation for this Letter
in which we test the {\it WMAP} data for SI, in detail, using
the  bipolar power spectrum (BiPS) of CMB anisotropy, a powerful
and easy to understand statistics, which was recently introduced to
the field of CMB analysis \cite{us_apjl, us_pascos}.  If SI is violated, the $C_l$ do not contain all the statistical
information of a Gaussian sky map. It is described by the whole set of
cross-correlations of the coefficients in the harmonic space $\langle
a_{lm} a_{l'm'}\rangle$. The cosmic variance of the elements will
prevent us from estimating the individual elements from a single map.
Hence, it is very important to define a statistic that condenses the
information of that matrix into a single spectrum that reduces the
cosmic variance. BiPS does exactly that. Furthermore it is easy to
interpret and straightforward to implement for very rapid computation.

\section{Bipolar Power Spectrum  Analysis} 

In spherical harmonic space representation where $\Delta T(\hat n)=
\sum_{lm}a_{lm}Y_{lm}(\hat n)$, the condition for SI
translates to a diagonal $\langle a_{lm}a^*_{l^\prime
m^\prime}\rangle=C_{l}\delta_{ll^\prime} \delta_{mm^\prime}$ where
$C_l$, 
is a complete description of (Gaussian) CMB anisotropy. When $\langle
a_{lm} a^*_{l^\prime m^\prime}\rangle$ is not diagonal, the off-diagonal 
elements contain information beyond $C_l$.

The BiPS is a combination of these
off-diagonal elements in $\langle a_{lm} a^*_{l^\prime
m^\prime}\rangle$, containing extra non-SI
information. The BiPS together with $C_l$ give a complete,
orientation independent description of a Gaussian CMB anisotropy map.

The BiPS of CMB is defined by~\cite{us_apjl,us_pascos,us_prd}
\begin{equation}\label{klharm}
\kappa_\ell =\sum_{ll^\prime M} |A_{ll^\prime}^{\ell M}|^2 \geq 0\,,
\end{equation}
where $A_{ll^\prime}^{\ell M}$ are the coefficients of the bipolar
spherical harmonic (BiPoSH) expansion of the correlation function, which is the most general expansion of $C(\hat{q},\, \hat{q}^\prime)$:
\begin{equation}\label{bipolar}
 C(\hat{q},\, \hat{q}^\prime)\, =\, \sum_{ll^\prime\ell M} A_{ll^\prime}^{\ell
  M}\{Y_{l}(\hat{q}) \otimes Y_{l^\prime}(\hat{q}^\prime)\}_{\ell M}\,.
\end{equation}
BiPoSH functions are the natural complete basis on $S^2 \times S^2$ \cite{Var}.
Here $A_{ll^\prime}^{\ell M}$
are a complete set of linear combinations of off-diagonal elements of $\langle
a_{lm}a^*_{l^\prime m^\prime}\rangle$:
\begin{equation}\label{alml1l2}
 A_{ll^\prime}^{\ell M} = 
\sum_{m m^\prime} \langle a_{lm}a^*_{l^\prime m^\prime}\rangle
\, \, (-1)^{m^\prime} \mathfrak{ C}^{\ell M}_{lml^\prime -m^\prime}\,,
\end{equation}
where $\mathfrak{C}^{\ell M}_{lml^\prime m^\prime}$ are Clebsch-Gordan
coefficients.  Every specific measure of off-diagonal elements of the covariance 
matrix $\langle a_{lm}a^*_{l^\prime m^\prime}\rangle$ suggested in the 
literature \cite{DKY, prunet} can be expressed  in terms of  $A_{ll^\prime}^{\ell M}$.
For an SI correlation function $\langle a_{lm}a^*_{l^\prime m^\prime}
\rangle = C_{l} \delta_{ll^\prime}\delta_{mm^\prime}$, implying
$A_{ll^\prime}^{\ell M}=(-1)^l C_{l} (2l+1)^{1/2} \,
\delta_{ll^\prime}\, \delta_{\ell 0}\, \delta_{M0}$ and 
$\kappa_\ell \, = \, \kappa^0 \delta_{\ell 0}$.  An equivalent real
space description of the BiPS  in terms of appropriately weighted averages of
the correlation $C(\hat{q},\hat{q}^\prime)$  is given
in previous publications \cite{us_apjl, us_pascos}.

We compute the BiPS 
using fast methods of spherical harmonic transform of the
map~\footnote{We use the ANAFAST routine of the HEALPix
package~\cite{hpix} publicly available at
http://www.eso.org/science/healpix/.} and apply filters in multipole space using
the appropriate positive definite window function $W_l$. We define an
unbiased estimator for the BiPoSH coefficients  and then estimate 
$\kappa_\ell$, 

\begin{equation}\label{klALMesthar}
\tilde A_{ll^\prime}^{\ell M} = \sqrt{W_l W_{l^\prime}} \, \sum_{m
m^\prime} a_{lm}a_{l^\prime m^\prime} \, \, \mathfrak{ C}^{\ell
M}_{lml^\prime m^\prime}\,,\quad\quad \tilde\kappa_\ell =\sum_{ll^\prime M}
\left|\tilde A_{ll^\prime}^{\ell M}\right|^2 - {\mathfrak B}_\ell\,,
\end{equation}
where  ${\mathfrak B}_\ell \equiv\langle\tilde\kappa_\ell^B\rangle_{_{\rm SI}}$, is the bias that arises from the SI part of the correlation function. Although BiPS is quartic in $a_{lm}$, it is designed to detect SI violation and not non-Gaussianity. 

Consequently, for SI correlation, the measured $\tilde \kappa_\ell$
will be consistent with zero within the cosmic variance. Cosmic error, 
$\sigma_{_{\rm SI}}$, and bias, ${\mathfrak B}_\ell$, are given by equations (14) and (17) of \cite{us_apjl} respectively, with
$C_{l} \to W_lC_l$, where $C_l = C_l^S + C_l^N$ is the total SI angular power spectrum of the ``true'' signal and noise. Multipole space windows that weigh down the contribution from the rest of the SI region of the multipole space will  enhance the signal relative to cosmic error, $\sigma_{_{\rm SI}}$. 

 It is important to note that bias can never be subtracted exactly
for a non-SI map. What is important is whether the measured $\tilde
\kappa_\ell$ differs from zero at a statistically significant
level.

It is not inconceivable that for strong SI
violation, ${\mathfrak B}_\ell$ over-corrects for the bias leading to
negative values of $\tilde \kappa_\ell$.   The  noise covariance
is a possible source of SI violation. However, we have restricted
our analysis to $l \lsim 60$ where the errors are dominated by the cosmic variance.

\begin{figure*}[h]
  \includegraphics[scale=0.3, angle=-90]{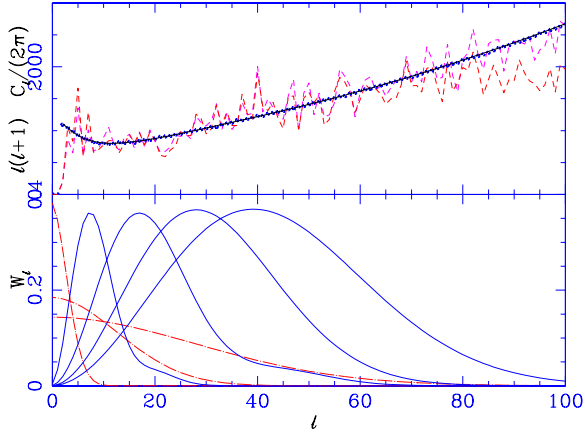}
 \caption{  {\em Top:} 
$C_{\ell}$ of the two {\it WMAP} CMB anisotropy maps. The red and magenta curves correspond to map A and B, respectively. The black line is a ``best-fit'' {\it WMAP} theoretical $C_{\ell}$ used for simulating SI maps. Blue dots are the average
$C_l$ recovered from $1000$ realizations. {\em Bottom:} These plots show
 the window functions used. The dashed
curves with increasing $l$ coverage are ``low-pass'' filter, $W_l^G(l_s)$, with $l_s=4, 18, 40$, 
respectively.  The solid lines are ``band-pass'' filter $W^S_l(l_t,l_s)$ with 
$(l_s,l_t)=(13,2), (30,5), (30,20), (45,20)$, respectively.}
\label{cl_wmap_cl100}
\end{figure*}

We carry out measurement of the BiPS, $\kappa_\ell$, on two full sky
CMB anisotropy maps -- (A) the Internal Linear Combination map (denoted
as ``ILC'' in the figures; \cite{wmap}), and (B) a foreground cleaned map
(denoted as ``Tegmark''; \cite{maxwmap}).  The angular power spectra of
these maps, shown in Figure \ref{cl_wmap_cl100}, are consistent with the best fit
theoretical power spectrum from the {\it WMAP} analysis.~\footnote{Based on
an LCDM model with a scale-dependent (running) spectral index.} 

We use simple filter functions in $l$
space to isolate different ranges of angular scales. A low pass,
Gaussian filter
\begin{equation}
W^G_l(l_s) = \exp[-(l+1/2)^2/(l_s+1/2)^2],
\label{gaussfilter}
\end{equation}
which cuts off power on small angular scales ($\lsim 1/l_s$), 
and a band pass filter,
\begin{equation}
W^S_l(l_t, l_s) = (2\{1- J_0[(l+1/2)/(l_t+1/2)]
\})\,W^G_l(l_s)
\label{bpfilter}
\end{equation}
which retains power within a range of multipoles set by $l_t$ and $l_s$.  
The window functions used in our work are plotted in
Figure \ref{cl_wmap_cl100}.

We use the {\it WMAP} best-fit (WMAPbf) $C_l$ to generate 1000 simulations of the SI CMB
maps. The $a_{lm}$ are generated up to an $l$ of 1024
(corresponding to maps at HEALPix resolution $N_{side}=512$). These are then
multiplied by the window functions $W^G_l(l_s)$ and $W^S_l(l_t,
l_s)$ and the BiPS for each realization is computed. 
 We use $C_l^T$ to analytically compute the bias and the
cosmic variance for $\kappa_{\ell}$. The average and standard deviation of
$\kappa_{\ell}$ of the SI maps are also estimations of the bias and 
 the cosmic variance.
We verified that the theoretical cosmic variance and bias match the numerical estimations of standard deviation and
average $\kappa_{\ell}$ of the $1000$ realizations of the SI maps.

\begin{figure*}[h]

  \includegraphics[scale=0.3, angle=-90]{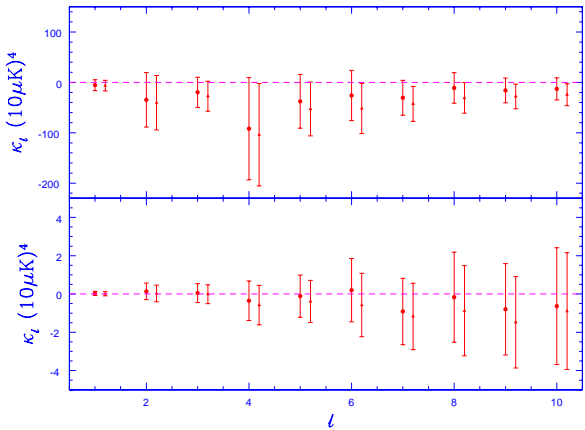}
  \caption{Measured values of $\kappa_\ell$ for maps
A and B filtered by $W_l^G(40)$ ({\it top panel}) and a Gaussian filter
$W_l^S(20,30)$ ({\it bottom panel}). The apparent non zero $\kappa_{\ell}$ 
at  
$\sim 1$ $\sigma$ is 
because the PDF is skewed 
 to negative values and also
partly because the {\it WMAP} theoretical spectrum has excess power at low
$l$ (see fig.~\ref{kappa_wmap_0_4}).}
\label{kappa_wmap_3020}
\label{kappa_wmap_400}
\end{figure*}

The BiPS of the full-sky CMB maps (A) and (B) based on the {\it WMAP} first-year
data is computed in the same manner.  We use the WMAPbf $C_l^T$ to
compute the bias and cosmic variance analytically. Using the
analytical bias and cosmic variance allows us to rapidly compute BiPS
with 1 $\sigma$ error bars for different theoretical $C_l^T$.
Figure~\ref{kappa_wmap_3020} shows the measured values of
$\kappa_\ell$ for maps (A) and (B) for two of the window functions.  We
compute the BiPS for all the window functions shown in
Figure \ref{cl_wmap_cl100}. 

\begin{figure*}[h]
 \includegraphics[scale=0.5, angle=0]{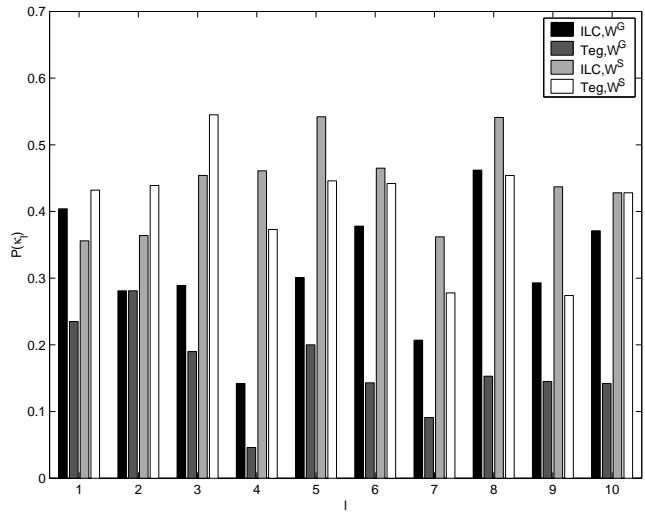}
  \caption{The probability of the two {\it WMAP} based CMB
  maps being  SI when filtered by $W_l^S(20,30)$ 
  and a Gaussian filter $W_l^G(40)$. } 
\label{prob3020}
\label{prob400}
\end{figure*}

The BiPS measured from the $1000$ simulated SI realizations of WMAPbf
$C_l$ is also used to estimate the probability distribution functions
(PDFs), defined as $p(\tilde \kappa_{\ell})=\int_{\tilde\kappa_l}^{\infty} d\kappa_l \,p(\kappa_l)$ for $\tilde\kappa_l > 0$ and $\int_{-\infty}^{\tilde\kappa_l}d\kappa_l \,p(\kappa_l)$ for $\tilde\kappa_l < 0$ (Fig.~\ref{prob3020}). 
We compute the individual probabilities of the map
being SI for each of the measured $\kappa_{\ell}$. 

 The probabilities for the $W^S_l(20,30)$
window function are greater than $0.25$ and the minimum probability at
$\sim 0.05$ occurs at $\kappa_4$ for $W_l^G(40)$. The reason for systematically lower SI probabilities for $W{_l}^S(20,30)$ as compared to $W{_l}^G(40)$ is simply due to the lower cosmic variance of the former. The contribution to the cosmic variance of BiPS is dominated by the low spherical harmonic multipoles. Filters that suppress the $a_{lm}$ at low multipoles have a
lower cosmic variance.

\begin{figure*}[h]
  \includegraphics[scale=0.3, angle=-90]{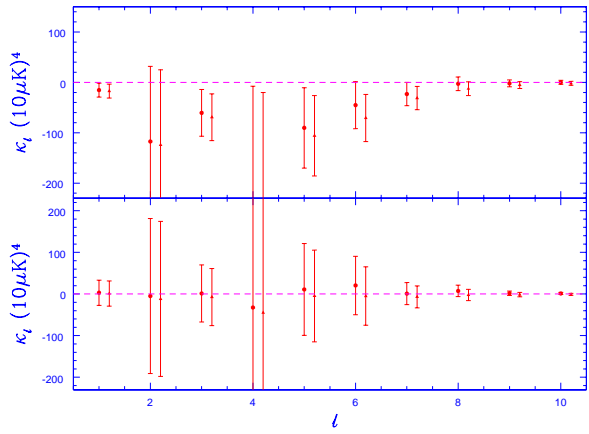}
  \caption{Measured values of BiPS for
maps A and B filtered to retain power only at
$l=2$ and $3$, assuming 
WMAPbf ({\it top})
and a model spectrum that matches the suppressed power at the lowest
multipoles \cite{shaf_sour04}.}
\label{kappa_wmap_0_4}
\end{figure*}

It is important to note that this probability is a conditional
probability of the measured $\tilde\kappa_{\ell}$ being SI given the
theoretical spectrum $C_l^T$ used to estimate the bias.  A final
probability emerges as the Bayesian chain product with the probability
of the theoretical $C_l^T$ used given data. Hence, small differences in
these conditional probabilities for the two maps are perhaps not
 significant. 
The important role played by the choice of the theoretical
model for the BiPS measurement is shown for a $W_l$ that retains power
in the lowest multipoles, $l=2$ and $3$. Assuming WMAPbf $C_l^T$,
there are hints of non-SI detections in the low $\ell$ (Fig. \ref{kappa_wmap_0_4}, {\it top}). We also compute the BiPS using a
$C_l^T$ for a model that accounts for the suppressed quadrupole and
octopole in the {\it WMAP} data \cite{shaf_sour04}. The mild detections of a
non zero BiPS vanish for this case (Fig.~\ref{kappa_wmap_0_4}, {\it bottom}).  
This suggests that the excess power in the WMAPbf $C_l^T$ with respect 
to the measured $C_l$ from {\it WMAP} at the lowest multipoles tends to indicate mild deviations from SI.

The SI of the CMB anisotropy has been under
scrutiny after the release of the first year of {\it WMAP} data. 
 Using a BiPS analysis we find no strong evidence for SI violation 
in the WMAP CMB anisotropy maps. We have verified that our null
results are consistent with measurements on simulated SI maps.   

It is also possible to construct a model-independent ``frequentist''
estimator of BiPS that uses the $C_l$ of the map itself to compute the
bias. Measurements on our simulated sky maps show that cosmic variance for
this estimator is much smaller. The preliminary results on the same set
of observed maps are consistent
with SI and will be reported later. Work is in progress to verify
them against analytical results. The full-sky maps and the restriction to
low $l \lsim 60$ (where instrumental noise is sub-dominant) permits the use
of our analytical bias subtraction and error estimates. The excellent
match with the results from numerical simulations is a strong
verification of the numerical technique.  
This is an important check before using Monte-Carlo simulations in future work targeting the $l\gsim 100$ regime, which involves a galactic mask and a non-uniform noise matrix.

There are strong theoretical motivations for hunting for SI
violation in the CMB anisotropy. The possibility of non-trivial cosmic
topology is a theoretically well-motivated possibility that has also
been observationally targeted \cite{ell71, lev02}.  
Violation of SI is a generic feature of 
cosmic topology \cite{bps}.
And the BiPS expected in flat,
toroidal models of the universe has been computed and shown to be
related to the principle directions in the Dirichlet
domain \cite{us_prl}. 
Hence, the null result of BiPS have important implications for
cosmic topology. This approach complements the direct search for the signature
of cosmic topology,  and our results are consistent
with the absence of the matched circles and the null S-map test of the
{\it WMAP} CMB maps \cite{circles04, angelica}. 
Work is in
progress to combine a full Bayesian likelihood analysis with 
 BiPS analysis to constrain cosmic topology \cite{us_dodeca}, 
in particular the recently proposed dodecahedron universe \cite{lum03}. 
We defer to future publication, detailed analysis and constraints on
cosmic topology using null BiPS measurements. Other
theoretical scenarios that predict the breakdown of SI are also being
probed using BiPS, e.g., primordial
cosmological magnetic fields \cite{DKY, gang}.

Observational artifacts such as foreground residuals, non-circular
beam, inhomogeneous noise correlation, residual stripping patterns,
etc.  are potential sources of SI breakdown.  Our null BiPS results
confirm that these artifacts do not significantly contribute to the maps
studied here.

\section{Summary and Conclusion}
In summary, we find null measurements of the BiPS for a selection of
full-sky CMB anisotropy maps based on the first year of {\it WMAP} data. Our
results rule out radical violation of statistical isotropy, and are
consistent with null results for matched circles and the S-map tests
of SI violation.  We find that the excess power at 
low $l$ in the best-fit theory $C_l$ with respect to $C_l$ derived from 
the {\it WMAP} maps tends to enhance the SI violation signatures in BiPS. 
Our result is an unambiguous, well-defined,
quantitative evaluation and assessment of the SI of
the CMB anisotropy.  But pending a more careful comparison, the
results do not necessarily conflict with a number of other statistical
tests that have suggested violation of statistical isotropy. We should also mention that our null BiPS results do not have any bearing on the non-Gaussianity of the maps.


The authors are very thankful for close interaction with David
Spergel and Glenn Starkman all through the project, including careful
reading and valuable comments on previous drafts of this Letter. We also
acknowledge useful discussions with Dick Bond, Dmitry Pogosyan, and
Carlo Contaldi.  A.H. acknowledges help
from A. Bhattacharjee. The use of the HPC facility of IUCAA 
computation is acknowledged.


\end{document}